\documentclass[cernpreprint, backref=false, texlive=2020, UKenglish, dvipsnames, block=none, texmf]{atlasdoc}
 
\usepackage{hyperref}
\usepackage{atlaspackage}
\usepackage{atlasbiblatex}
\usepackage{atlasphysics}
\graphicspath{{logos/}{figures/}}

\AtlasTitle{Dijet resonance search with weak supervision \\ using $\sqrt{s}=13$ TeV $pp$ collisions in the ATLAS detector}

\AtlasAbstract{
This Letter describes a search for narrowly resonant new physics using a machine-learning anomaly detection procedure that does not rely on a signal simulations for developing the analysis selection.  Weakly supervised learning is used to train classifiers directly on data to enhance potential signals.  The targeted topology is dijet events and the features used for machine learning are the masses of the two jets.  The resulting analysis is essentially a three-dimensional search $A\rightarrow BC$, for $m_A\sim\mathcal{O}(\text{TeV})$, $m_B,m_C\sim\mathcal{O}(100\text{ GeV})$ and $B,C$ are reconstructed as large-radius jets, without paying a penalty associated with a large trials factor in the scan of the masses of the two jets.  The full Run 2 $\sqrt{s}=13$ TeV $pp$ collision data set of 139 fb$^{-1}$ recorded by the ATLAS detector at the Large Hadron Collider is used for the search.   There is no significant evidence of a localized excess in the dijet invariant mass spectrum between 1.8 and 8.2 TeV.   Cross-section limits for narrow-width $A$, $B$, and $C$ particles vary with $m_A$, $m_B$, and $m_C$.  For example, when $m_A=3$ TeV and $m_B\gtrsim 200$ GeV, a production cross section between 1 and 5~fb is excluded at 95\% confidence level, depending on $m_C$.  For certain masses, these limits are up to 10 times more sensitive than those obtained by the inclusive dijet search.  These results are complementary to the dedicated searches for the case that $B$ and $C$ are Standard Model bosons.
}
 
\AtlasJournalRef{Phys. Rev. Lett. 125 (2020) 131801}
\AtlasDOI{10.1103/PhysRevLett.125.131801}
 
\PreprintIdNumber{CERN-EP-2020-062}
 
\author{The ATLAS Collaboration}
 
\AtlasRefCode{HDBS-2018-59}

\hypersetup{pdftitle={ATLAS document},pdfauthor={The ATLAS Collaboration}}
 
\begin{document}
 
\maketitle

A search for dijet resonances is one of the first analyses performed when a hadron collider reaches a new center-of-mass energy~\cite{Albajar:1988rs,Alitti:1990kw,Abe:1989gz,Abazov:2003tj,Aad:2010ae,Khachatryan:2010jd,Khachatryan:2015dcf,ATLAS:2015nsi}.   While such searches are sensitive to nearly all resonance decays $A\rightarrow BC$, dedicated searches for particular decays will always be more sensitive.  This is the motivation for dedicated resonance searches for the case where $B$ and $C$ are $\tau$-leptons~\cite{Aaboud:2017sjh,Khachatryan:2016qkc}, $b$-quarks~\cite{Aaboud:2019zxd,Aaboud:2018tqo,Sirunyan:2018ikr}, top quarks~\cite{Aaboud:2019roo,Sirunyan:2018ryr}, vector bosons~\cite{Aaboud:2018bun,Sirunyan:2019jbg}, Higgs bosons~\cite{Aaboud:2018sfw,Aaboud:2018knk,Sirunyan:2019quj,Aaboud:2018zhh,Sirunyan:2018qca,Sirunyan:2018fuh}, and more, including asymmetric combinations. In all cases, a selection on the structure of the energy flow from each side of the decay is used to enhance events with the targeted topology.  Searches for any combination of Standard Model (SM) particles can be well-motivated by one or more theory frameworks beyond the SM (BSM), but not all combinations are currently covered by dedicated searches~\cite{Craig:2016rqv}.  Furthermore, there are only a small number of searches~\cite{Aaboud:2018eoy,Aaboud:2017ecz,Aaboud:2018djx,Aaboud:2018aqj,Aaboud:2019opc,Sirunyan:2018vhk,Sirunyan:2017yrk,Sirunyan:2018fki,Sirunyan:2018rfo,Aaboud:2018zba,Aaboud:2018fzt,Aaboud:2018zba,Sirunyan:2019vxa,Sirunyan:2017dnz,Sirunyan:2017nvi,Sirunyan:2019sgo} that cover the vast set of possibilities where at least one of $B$ or $C$ is itself a BSM particle~\cite{Kim:2019rhy}.  There is no previous search where all of $A,B$ and $C$ are BSM particles and can have different masses.
 
While it is crucial to continue searching for particular dijet topologies, the fact that not all SM and BSM possibilities are covered suggests that a complementary generic search effort is required.  What is needed is a method for searching for many topologies all at once that ideally does not pay a large statistical trials factor.   A variety of existing and proposed model-agnostic searches range from nearly signal model-independent but fully background model-dependent~\cite{sleuth,Abbott:2000fb,Abbott:2000gx,Abbott:2001ke,Aaron:2008aa,Aktas:2004pz,Aaltonen:2007dg,Aaltonen:2007ab,Aaltonen:2008vt,CMS-PAS-EXO-14-016,CMS-PAS-EXO-10-021,Aaboud:2018ufy,ATLAS-CONF-2014-006,ATLAS-CONF-2012-107,DAgnolo:2018cun,DAgnolo:2019vbw} (because they compare data with SM simulation) to varying degrees of partial signal-model and background-model independence~\cite{Nachman:2020lpy,Andreassen:2020nkr,Farina:2018fyg,Heimel:2018mkt,Roy:2019jae,Cerri:2018anq,Blance:2019ibf,Hajer:2018kqm,DeSimone:2018efk,Mullin:2019mmh,1809.02977,Dillon:2019cqt,Aguilar-Saavedra:2017rzt,Collins:2018epr,Collins:2019jip,Amram:2020ykb}.  The method used for this analysis employs a machine-learning-based anomaly detection procedure to perform a dijet search in which the jets from a potential signal have a nontrivial but unknown structure~\cite{Collins:2018epr,Collins:2019jip}.  Simply stated, classifiers are trained to distinguish particular dijet invariant mass bins from their neighbors.  Localized resonances will be enhanced with a selection based on the classifier.
 
This Letter presents a search for a generic $A\rightarrow BC$ resonance, in which all of $A,B$ and $C$ could be BSM particles and $m_B,m_C\ll m_A$ so that the decay products of $B$ and $C$ can be contained within single large-radius jets.  A search by the CMS collaboration~\cite{Sirunyan:2019jbg} involves a three-dimensional fit over jet and dijet masses, but results are reported only for the case that $B$ and $C$ are $W$ or $Z$ bosons.  The analysis presented here achieves sensitivity to BSM cases without performing a scan in masses other than from the $A$ particle.  In particular, the search uses events collected by the ATLAS detector~\cite{PERF-2007-01,PIX-2018-001} using the full $139$~\ifb~Run 2 $\sqrt{s}=13$~\TeV\ $pp$ collision data set.   Weakly supervised classifiers are used to enhance potential signals without using simulations of any particular signal models.
 
Events with at least two jets are considered, and the invariant mass distribution of the two leading jets is used to perform a `bump hunt'.  Jets are formed~\cite{Cacciari:2011ma,Cacciari:2005hq} from locally calibrated calorimeter cell-clusters~\cite{Aad:2016upy} using the anti-$k_t$ algorithm~\cite{Cacciari:2008gp} with a radius parameter of $R=1.0$.  These jets are trimmed~\cite{Krohn:2009th} by reclustering the jet constituents with the $k_t$ algorithm using $R=0.2$ and removing the constituents with transverse momentum ($p_\text{T}$) less than $5\%$ of the original jet $p_\text{T}$.  The jet four-vectors are then calibrated as detailed in Ref.~\cite{Aaboud:2018kfi}.
The two jets are required to each have $p_\text{T}>200$~\GeV\ and pseudorapidity\footnote{Pseudorapidity is defined in terms of the angle $\theta$ relative to the beam line as $\eta=-\ln\tan(\theta/2)$.} $|\eta|<2.0$.
In order to be broadly sensitive to hadronically decaying narrowly resonant particles, events are required to have at least one jet with $p_\text{T}>500$~\GeV\ and two leading jets with a rapidity difference of $|\Delta y_\text{JJ}| < 1.2$.   The $p_\text{T}$ threshold is chosen so that the online trigger system is fully efficient~\cite{ATLAS:2016qun,Aaboud:2016leb}.   Furthermore, both jets must have jet mass $30~\GeV<m_\text{J}<500~\GeV$ for stability of the neural network (NN) training described below.   The upper threshold reduces the $m_\text{JJ}$-dependence of the $m_\text{J}$  distribution.  The bump hunt is performed for dijet invariant masses in the range $2.28~\TeV<m_\text{JJ}< 6.81~\TeV$.
 
The masses of the two leading jets are used for classification.
As the first application of fully data-driven machine-learning anomaly detection, this restricted feature set is used to establish the procedure and is already sensitive to a wide range of BSM possibilities.
Weakly supervised classifiers for high-energy physics~\cite{Metodiev:2017vrx,Komiske:2018oaa,Dery:2017fap,Cohen:2017exh} aim to distinguish signal from background without having labeled examples for training.
In particular, the classification without labels method~\cite{Metodiev:2017vrx} calls for two mixed samples that are statistically identical aside from different class proportions.
For this search, the two samples are constructed using signal regions in $m_\text{JJ}$ with width $20\%\times m_\text{JJ}$, chosen to correspond to the detector resolution for a narrow resonance.  The signal regions are labeled 0--7 and have boundaries: $[1.90, 2.28, 2.74, 3.28, 3.94, 4.73, 5.68, 6.81, 8.17]$~\TeV.   The last bin edge is chosen so that there are sufficiently many events for training targeting the penultimate region.
 
The jet mass probability density varies slowly with $m_\text{JJ}$, so neighboring regions in $m_\text{JJ}$ can be used to construct the mixed event samples required for weak supervision.
In particular, a network is trained to distinguish between a given $m_\text{JJ}$ \textit{signal} region and the two neighboring \textit{sideband} regions.  For the case in which some signal is present in the signal region, the network will learn to tag that signal and enhance a bump in the $m_\text{JJ}$ spectrum, while for the case in which there is no signal in the signal region, the tagging of the network will be essentially random, and the $m_\text{JJ}$ spectrum will remain smooth after tagging.
Since every signal region requires two neighboring sidebands, the $m_\text{JJ}$ regions 1--6 are chosen, and the entire process outlined below is repeated for each signal region.
In order to reduce existing correlations between jet mass and $m_\text{JJ}$,\footnote{This is similar to the method used in Ref.~\cite{Moult:2017okx}; additional decorrelation techniques are described in Refs.~\cite{Louppe:2016ylz,Dolen:2016kst,Stevens:2013dya,Shimmin:2017mfk,Bradshaw:2019ipy,ATL-PHYS-PUB-2018-014,DiscoFever,Xia:2018kgd,Englert:2018cfo,Wunsch:2019qbo}.} the jet masses $(m_1,m_2)$, with $m_1\geq m_2$, in each $m_\text{JJ}$ region are each mapped to be between $0$ and $1$.  This mapping is accomplished with the empirical cumulative distribution function of the marginal distribution over both jets, each of mass $m_\text{J}$: $m_\text{jet}\mapsto \frac{1}{n_\text{jets}}\sum_{i=1}^{n_\text{jets}}\mathbb{I}[m_\text{jet} > m_\text{jet $i$}]$.  The indictor function $\mathbb{I}[\cdot]$ is unity if its argument is true and zero otherwise, and the resulting marginal distribution is uniform in each $m_{\text{JJ}}$ bin (signal sensitivity is contained in the correlations).
Additional decorrelation is achieved by assigning the same total weight to each sideband.   Note that full decorrelation between the classifier and $m_\text{JJ}$ is sufficient but not necesary: sculpting an artificial bump requires that the classifier has at least a quadratic dependence on the dijet mass so that the efficiency is lower on the left side band, higher in the signal region, and then lower again in the right sideband.
The NNs performing the weakly supervised classification using $m_1$ and $m_2$ are fully connected networks built from four hidden layers with sizes 64, 32, 8, and 1.
Rectified linear units connect each intermediate hidden layer and the final activation function is sigmoidal.
Networks are implemented in Keras~\cite{keras} with the Tensorflow back end~\cite{tensorflow} and minimize the binary cross-entropy using the Adam optimizer~\cite{adam}.
 
In order to eliminate a trial factor associated with $(m_1,m_2)$, the NN identifies a region of interest, and no event is used to train the NN that is applied to it.
A $k$-fold cross-validation procedure is employed in which the full data set is divided randomly into $k$ parts of equal size.
Among these, $k-2$ parts are used for training $n$ classifiers (the training set) with different initializations, and the $(k-1)^\text{th}$ part is used to decide, based on the loss, which of these $n$ networks to select (the validation set).
The selected network is then mapped to an efficiency $\epsilon$ in the $k^\text{th}$ part (the test set) so that the meaning of the network output can be compared across data sets and trainings.  The efficiency $\epsilon$ is defined as the fraction of events with a given NN value or higher.
This output is averaged across the $k-1$ other permutations of the training and validation parts.
The entire procedure is then repeated $k$ times, where each part is a test set exactly once.
For this analysis, $k=5$ and $n=3$, so there are $3\times4\times 5=60$ NNs trained for each signal region.
Two event selections from thresholds imposed on the NN outputs are used: one that keeps the $10\%$ most signal-region-like events ($\epsilon=0.1$) and one that keeps the 1\% most signal-region-like events ($\epsilon=0.01$).
 
As the classifier-based event selection depends on the data, and in particular on the possible presence of true signals, it is not possible to directly define control regions to validate the method.  The entire procedure was validated using simulated events as well as a validation region with $|\Delta y_\text{JJ}|>1.2$.  For $s$-channel resonances, it is expected that this inverted rapidity-difference requirement reduces the signal efficiency while enhancing the dijet background by over an order of magnitude.  In these validation tests, the learning works effectively and there is no evidence for selection-induced excesses.  The expected limits are comparable to the ones that will be reported for the unblinded data in Fig.~\ref{fig:limits}.
 
\begin{figure}[h!]
\centering
\subfloat[]{\includegraphics[width=0.49\textwidth]{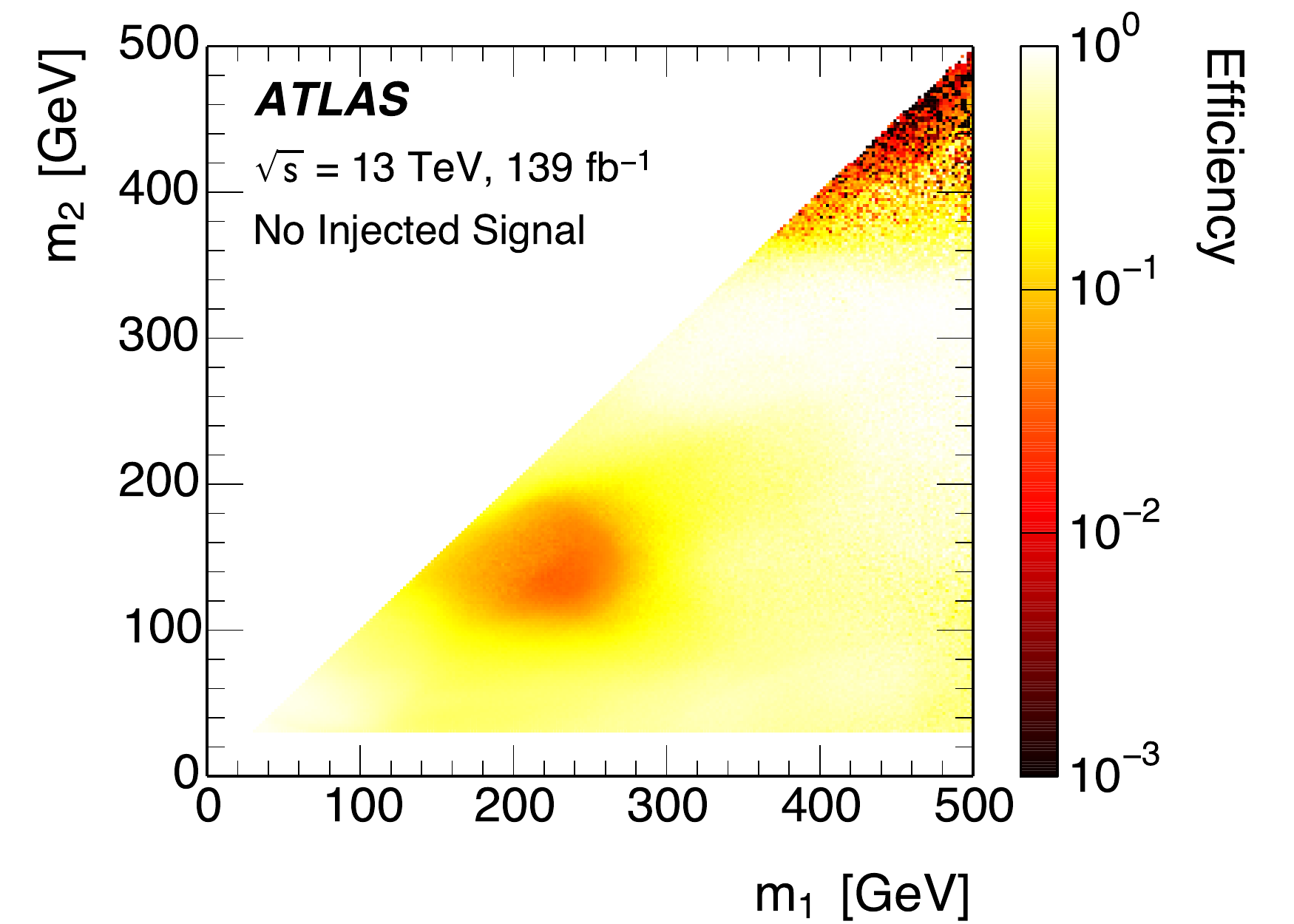}}
\subfloat[]{\includegraphics[width=0.49\textwidth]{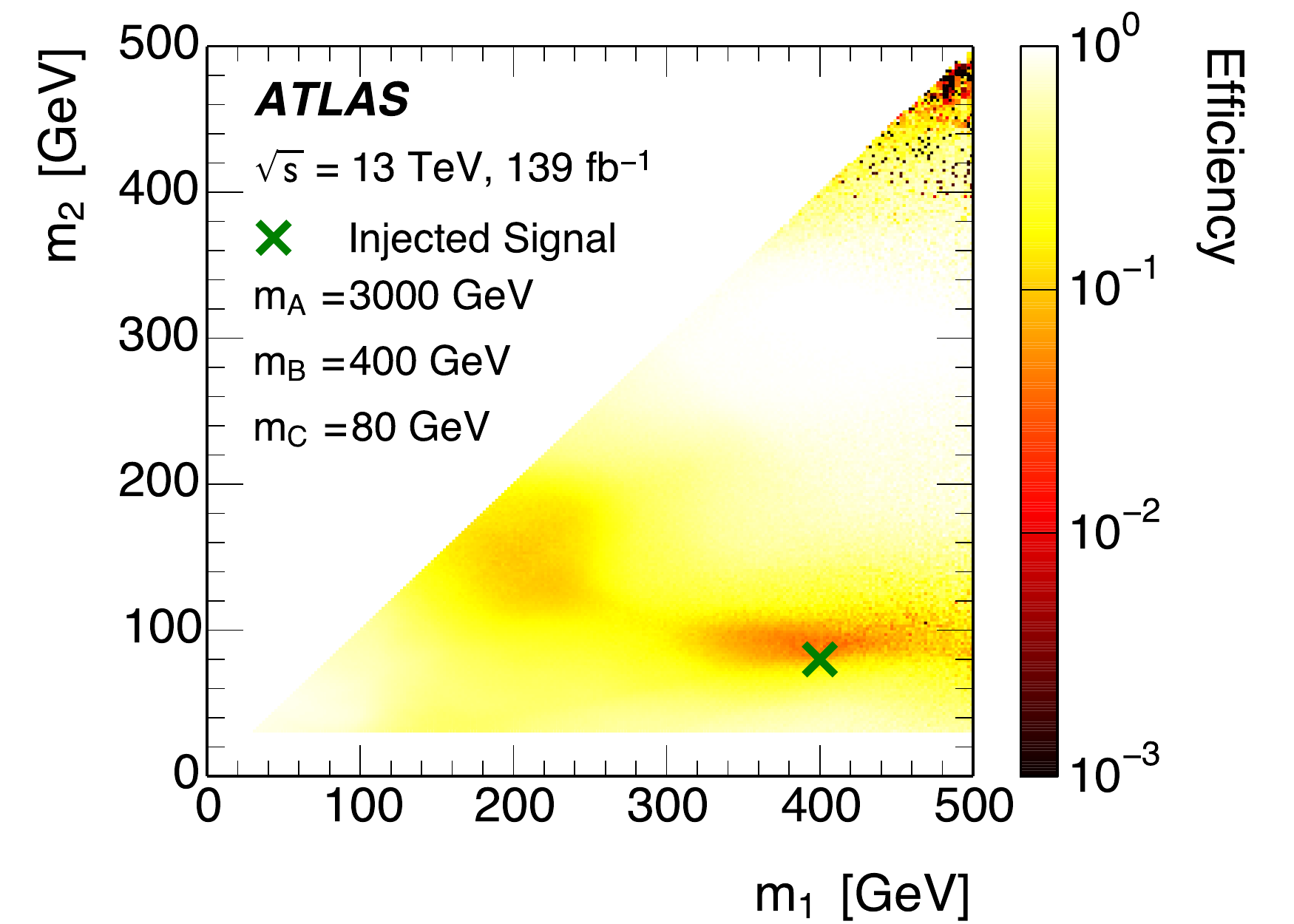}}\\
\subfloat[]{\includegraphics[width=0.49\textwidth]{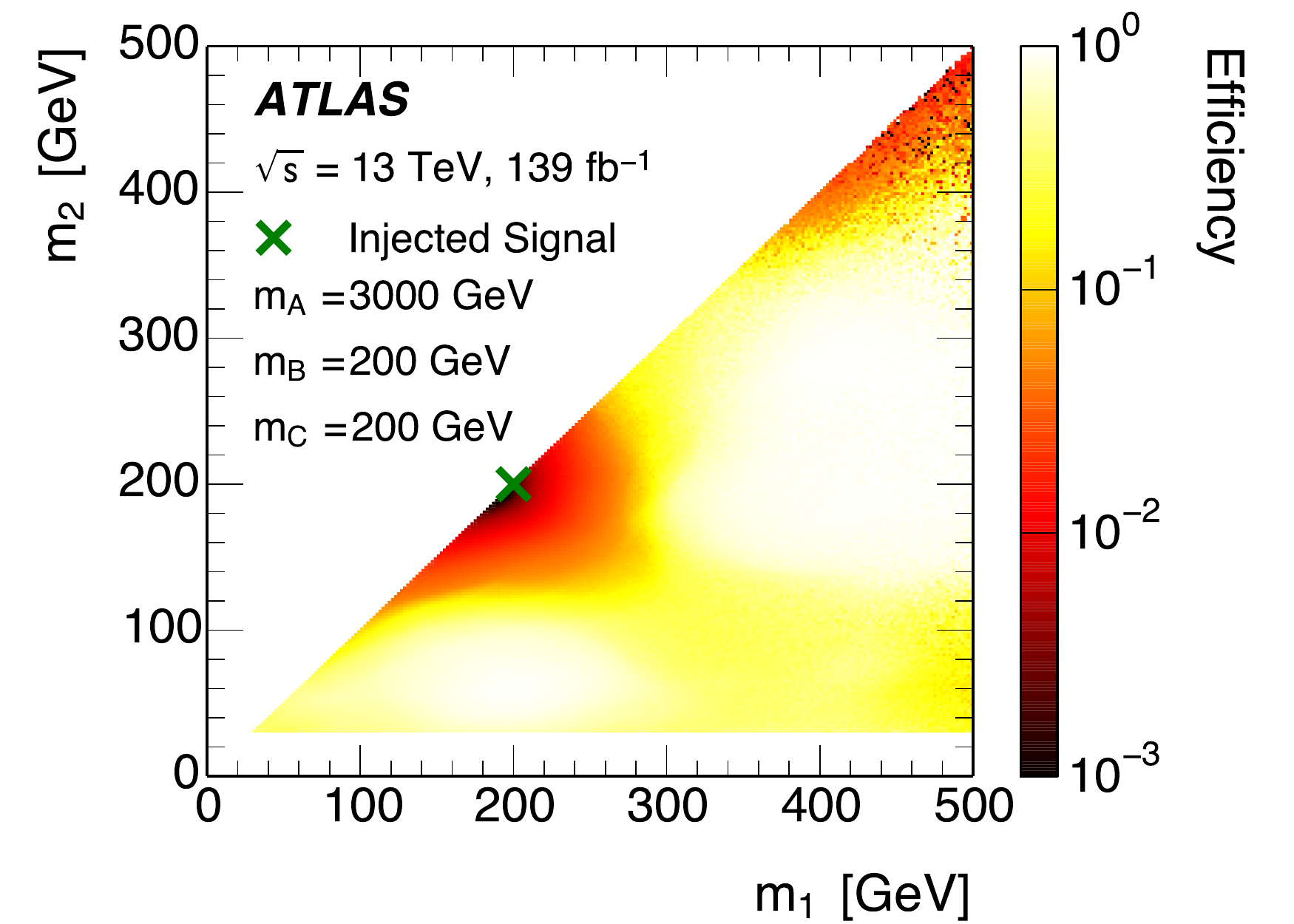}}
\subfloat[]{\includegraphics[width=0.49\textwidth]{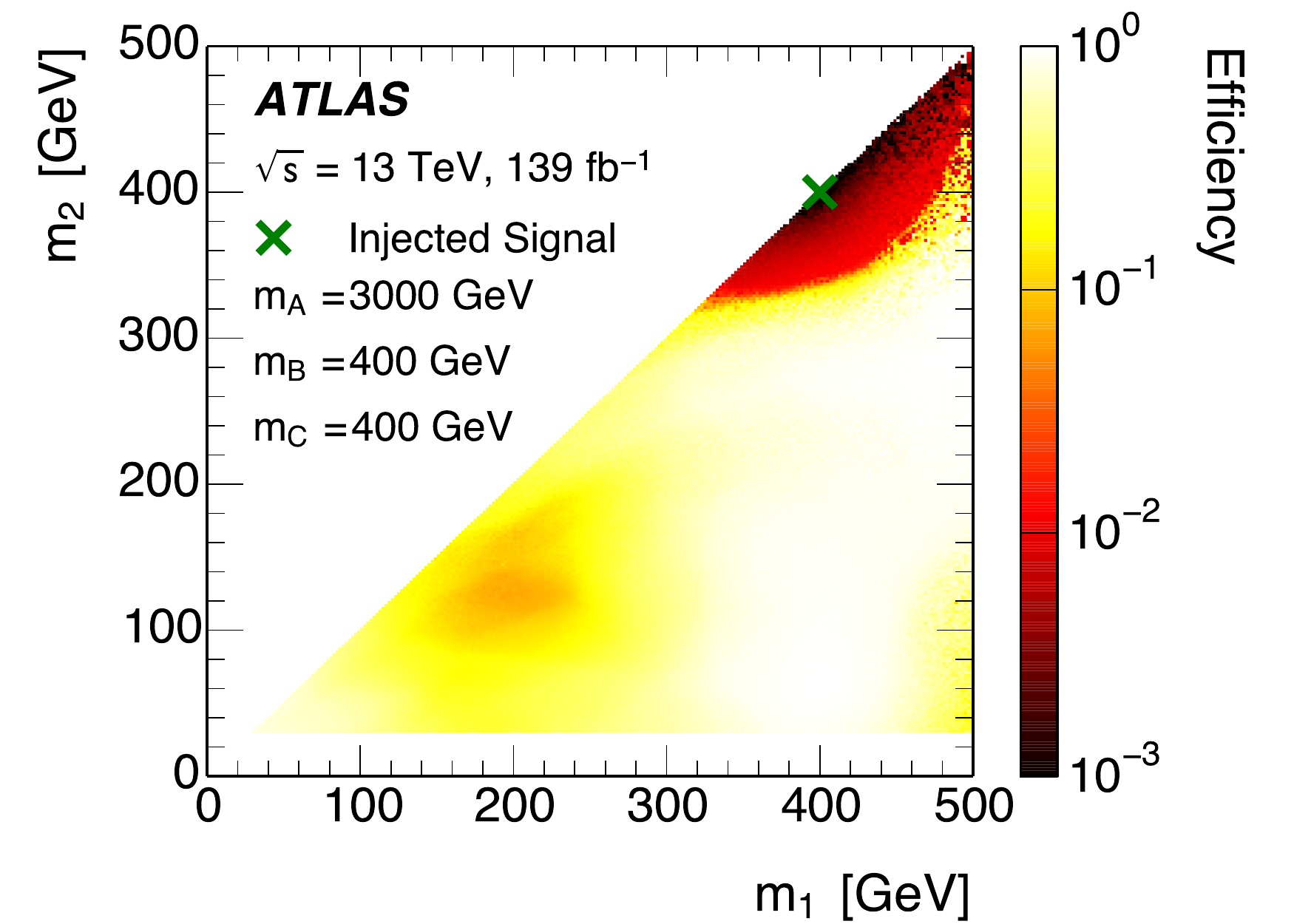}}
\caption{The efficiency mapped output of the NN versus the input variables for the events in signal region 2 for four cases: (a) there is no injected signal; (b) there is an injected signal of $m_A=3$~\TeV, and $m_B=400$~\GeV\ and $m_C=80$~\GeV, (c) there is an injected signal of $m_A=3$~\TeV, and $m_B=200$~\GeV\ and $m_C=200$~\GeV, and (d) there is an injected signal of $m_A=3$~\TeV, and $m_B=400$~\GeV\ and $m_C=400$~\GeV.  The location of $(m_B,m_C)$ for the given injected signal is marked with a green $\times$.  The injected cross section is just below the limit at low $m_B$ and $m_C$ from the inclusive dijet search~\cite{Aad:2019hjw}.  Additional signal region plots in the absence of an injected signal can be found at~\cite{aux}.
}
\label{fig:NNout}
\end{figure}
 
Following the validation, first, the performance of the NNs on data is studied with and without injected signals.  Since the NNs are two-dimensional functions, they can be visualized directly as images.
Figure~\ref{fig:NNout} presents the network output from a representative signal region in the absence of signal and also in the presence of injected signals.
By construction, there must be a region of low efficiency and the data are the same in all four plots.
In the absence of a signal, regions of low efficiency are located randomly throughout the $(m_1,m_2)$ plane.
The signals are $W'\rightarrow WZ$, for a new vector boson $W'$~\cite{Altarelli:1989ff}, and the $W$ and $Z$ boson masses are varied, with widths set close to zero.
These signals were simulated using \PYTHIAV{8.2}~\cite{Sjostrand:2014zea,Sjostrand:2007gs,Sjostrand:2006za} with the A14 set of tuned parameters~\cite{ATL-PHYS-PUB-2014-021} and NNPDF 2.3 parton distribution function~\cite{Ball:2012cx}.  All samples of simulated data were processed using the full ATLAS detector simulation~\cite{Aad:2010ah} based on $\GEANT4$~\cite{AGOSTINELLI2003250}.
The amount of signal injected in all cases is about the same as, or less than, the level already excluded by the all-inclusive dijet search~\cite{Aad:2019hjw}.
In all cases, the low-efficiency (signal-like) regions of the NN are localized near the injected signal.  Some signals are easier to find than others; the difficulty is set both by the relative size of the signal and by the total number of events available for training in the signal vicinity.

After applying an event selection based on the NN trained on a particular signal region, the $m_\text{JJ}$ spectra are fit with a parametric function.
The entire $m_\text{JJ}$ spectrum between 1.8 and 8.2~\TeV\ is fit with a binning of 100~\GeV; however, a \textit{fit signal region} and \textit{fit sideband region} are defined for evaluating the quality of the fit.
The fit signal regions are defined as the $m_\text{JJ}$ signal regions the NN used for training, combined with the adjacent halves of the left and right neighboring regions; the fit sidebands are defined as the complement of the fit signal regions.
An iterative procedure is applied until the $p$-value from the fit sideband $\chi^2$ is greater than 0.05.   Since the NN is trained to distinguish the signal region from its neighboring regions, it is expected that the $m_\text{JJ}$ spectrum is smooth in the fit sideband region in the presence or absence of a true signal.
First, the data are fit to $dn/dx=p_1(1-x)^{p_2-\xi_1 p_3}x^{-p_3}$, where $x=m_\text{JJ}/\sqrt{s}$, $p_i$ are fit parameters, and the $\xi_i$ are chosen to ensure that the $p_i$ are uncorrelated.
If the fit quality is insufficient, an extended function is used instead~\cite{Aad:2019hjw}: $dn/dx=p_1(1-x)^{p_2-\xi_1 p_3}x^{-p_3+(p_4-\xi_2p_3-\xi_3p_2)\log(x)}$.
If the fit quality remains insufficient, a variation of the UA2~\cite{Alitti:1990kw} fit function is tested: $dn/dx=p_1x^{p_2-\xi _3}e^{-p_3 x+(p_4-\xi_2 p_3-\xi_3p_2)x^2}$.
If the fit quality is still insufficient, the fit sidebands are reduced by $400$~\GeV\ on both sides and the three functions are tried again in order.
This procedure is then iterated until the fit is successful.
The fit results in the signal regions for the $\epsilon=0.1$ and $\epsilon=0.01$ NN efficiency selections are presented in Figure~\ref{fig:fits}.
The largest positive deviation from the fit model is $3.0\sigma$ in signal region $1$, around 2500~\GeV, at $\epsilon=0.1$ (the corresponding NN output does not show any significant features~\cite{aux}).
Globally, the positive tail of the signal region significance distribution is consistent with a standard normal distribution at the $1.5\sigma$ level.
 
\begin{figure}[h!]
\centering
\subfloat[]{\includegraphics[width=0.49\textwidth]{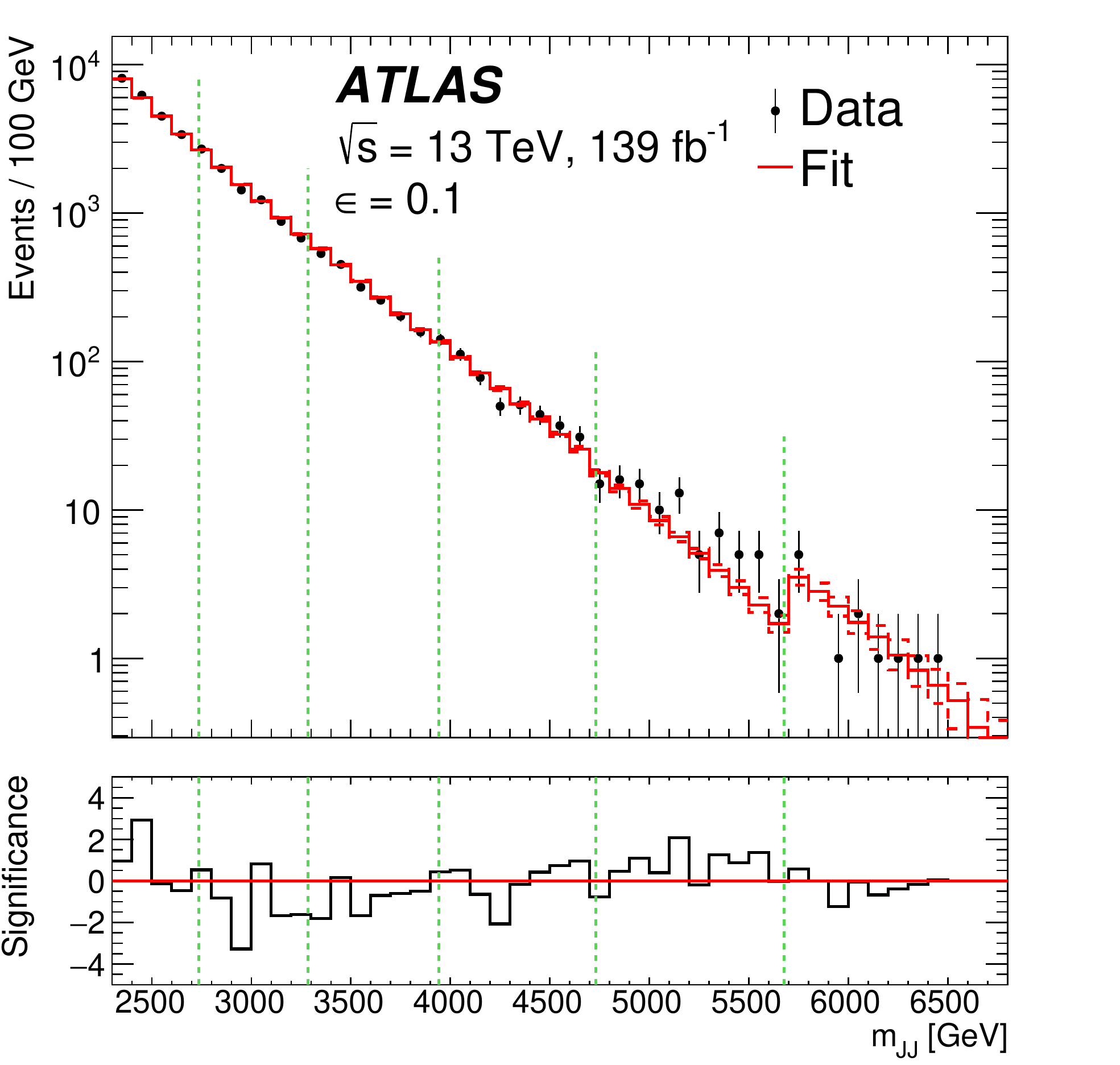}}
\subfloat[]{\includegraphics[width=0.49\textwidth]{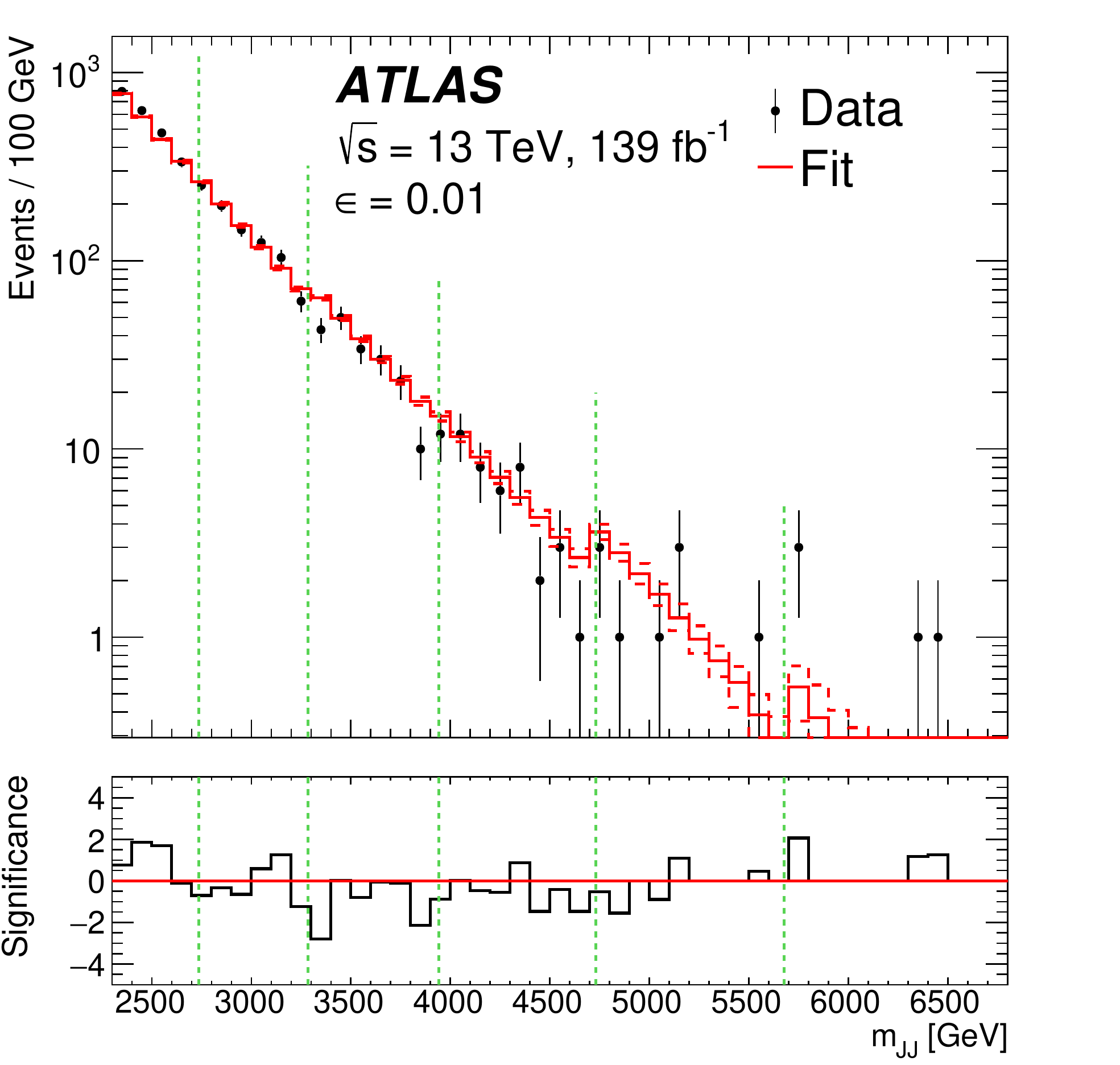}}\\
\subfloat[]{\includegraphics[width=0.49\textwidth]{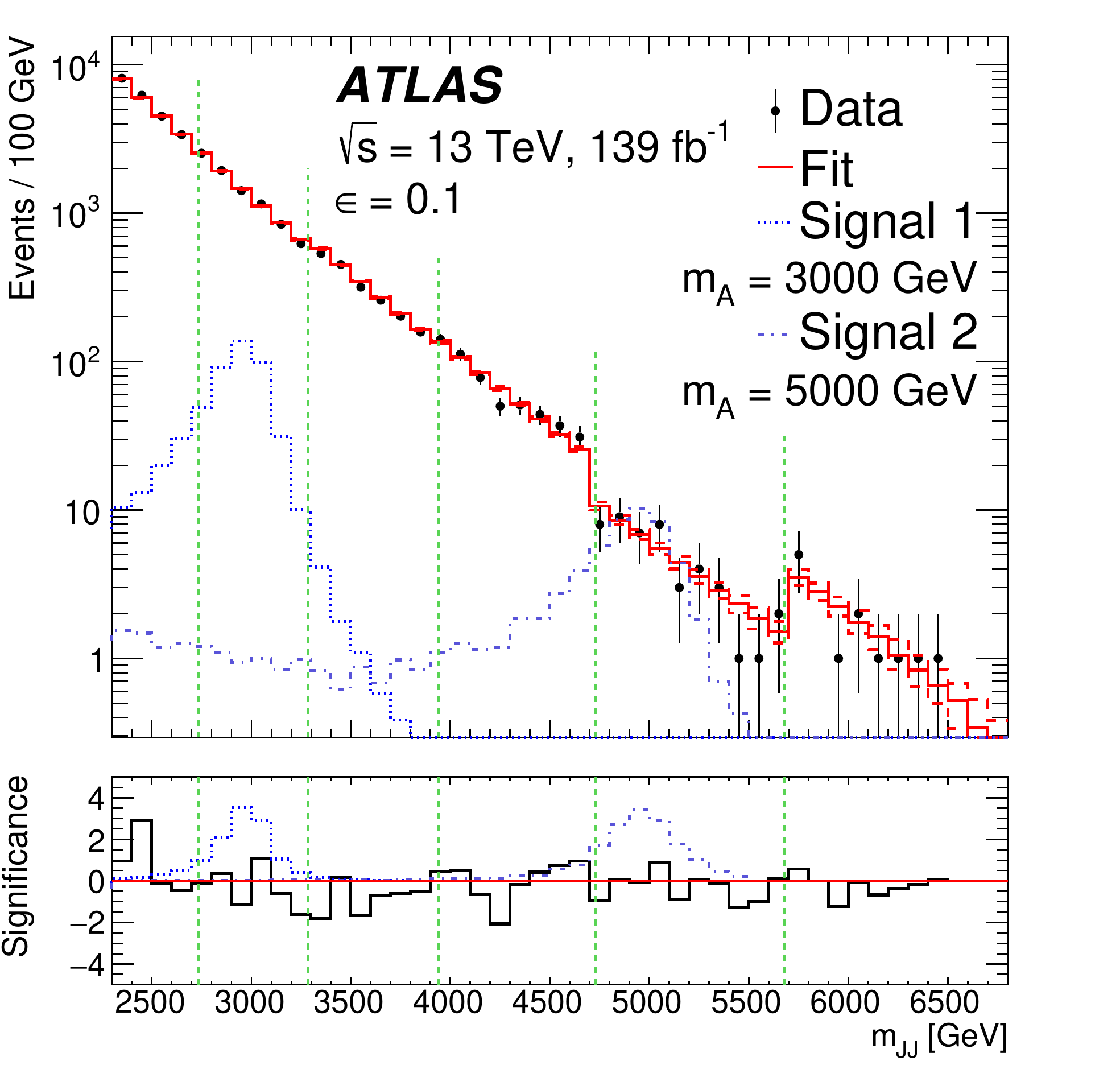}}
\subfloat[]{\includegraphics[width=0.49\textwidth]{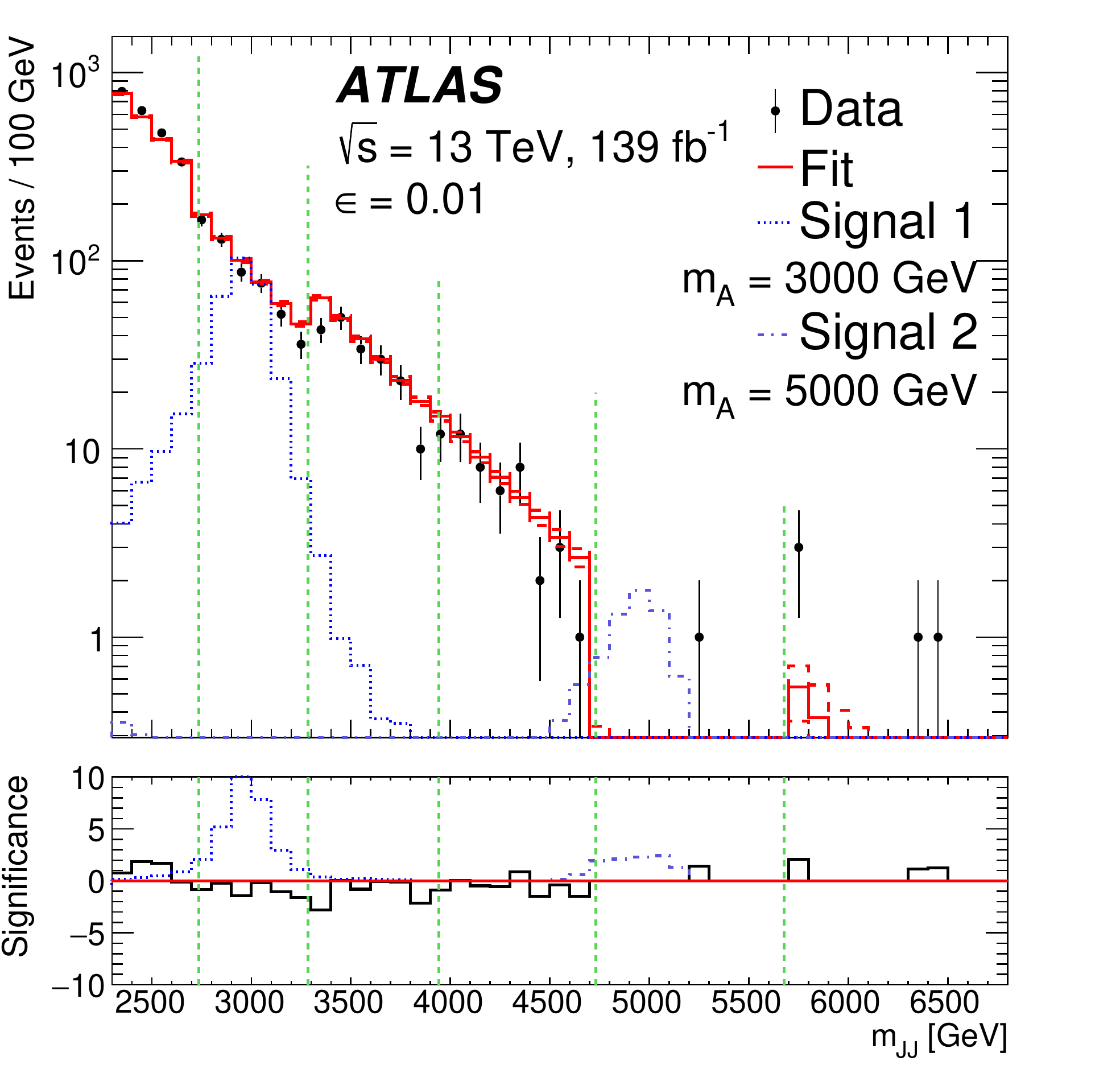}}
\caption{A comparison of the fitted background and the data in all six signal regions, indicated by vertical dashed lines, and for (a,c) $\epsilon=0.1$ and (b,d) $\epsilon=0.01$.  Dashed histograms represent the fit uncertainty.  The lower panel is the Gaussian-equivalent significance of the deviation between the fit and data.  The fits are performed including the sidebands, but only the signal region predictions and observations in each region are shown.  As the NN is different for each signal region, the presented spectrum is not necessarily smooth.  The top plots (a,b) show the result without injected signal, and the bottom plots (c,d) present the same results but with signals injected only for the NN training at $m_A=3$~\TeV~(Signal 1) and $m_A=5$~\TeV~(Signal 2), each with $m_B=m_C=200$~\GeV.  The injected cross section for each signal is just below the limit from the inclusive dijet search~\cite{Aad:2019hjw}.
}
\label{fig:fits}
\end{figure}
 
The $W'$ signal models can be used to set limits on the production cross section of specific new particles.
To illustrate the sensitivity of the analysis to the full three-dimensional parameter space $(m_A,m_B,m_C)$, two $m_A$ points and multiple $(m_B,m_C)$ points are selected.
As the NN performance depends on the data, the entire learning procedure has to be repeated every time a new signal model and signal cross section are injected into the data.
In order to reduce statistical fluctuations related to the shape of the signal, for each signal cross section the network is retrained with five random samplings from the signal simulation, and the network with the median performance is chosen.
A profile-likelihood-ratio test is used to determine 95\% confidence intervals for the excluded signal cross section.  When the number of expected events is much larger than one, asymptotic formulae~\cite{Cowan:2010js} are used for this test, otherwise, the test is performed numerically (only Fig.~\ref{fig:limits}d).  The excluded cross section is reported as $\max(\sigma_\text{CL},\sigma_\text{injected})$, where $\sigma_\text{CL}$ is the cross section determined from the profile-likelihood-ratio test and $\sigma_\text{injected}$ is the injected cross section.
This procedure is chosen because the network's performance may not be as good if there were truly less signal than was injected.
The resulting exclusion limits are presented in Figure~\ref{fig:limits}.
As the background expectation is determined entirely from data, the only systematic uncertainty associated with the background is the statistical uncertainty from the fit.
The only other relevant uncertainties are those related to the signal $m_{\text{JJ}}$ and $m_{\text{J}}$ modeling; experimental uncertainties in the reconstructed jet kinematics account for about a 10\% uncertainty in the excluded cross section.
 
The limits on $W'$ production vary with $m_A$, $m_B$, and $m_C$.
For $m_B=m_C=400$~\GeV, the excluded cross section is about 1~fb, a significant improvement over existing limits.
Lower $m_B$ and $m_C$ result in weaker limits because of the larger SM background in those regions; it is therefore difficult for the NN to learn to tag these signals.  The NN is most powerful when the local signal-to-background ratio is high and there are enough events for it to learn effectively.
For some models, such as $(m_A,m_B,m_C)=(5000,80,80)$~\GeV, the NN is not able to identify the signal effectively, resulting in limits weaker than those from previous searches.
For comparison, the sensitivities of the ATLAS inclusive dijet search (recast with signals from this paper)~\cite{Aad:2019fbh} and the all-hadronic diboson resonance search~\cite{Aad:2019hjw} are also shown in Figure~\ref{fig:limits}.
The inclusive dijet search sensitivity decreases for high $m_B$ and $m_C$ masses due to the use of small-radius jets that do not capture all of the $B$ and $C$ decay products.
The diboson resonance search has greater sensitivity when $m_B,m_C\approx m_W,m_Z$, but it has no sensitivity away from these points.  In this case, the diboson search uses more information than the weakly supervised one, but the trend is expected: assuming that the simulations used for developing the analysis selection are reliable, a fully supervised approach should outperform the weakly supervised one for any particular signal model.
Direct searches for $B$ and $C$ that trigger on initial-state radiation are also sensitive to these signal models~\cite{Aaboud:2018zba,Aaboud:2018fzt,Aaboud:2018zba,Sirunyan:2019vxa,Sirunyan:2017dnz,Sirunyan:2017nvi,Sirunyan:2019sgo}, but the sensitivity is much weaker than 10~fb.
\begin{figure}[h!]
\centering
\subfloat[]{\includegraphics[width=0.45\textwidth]{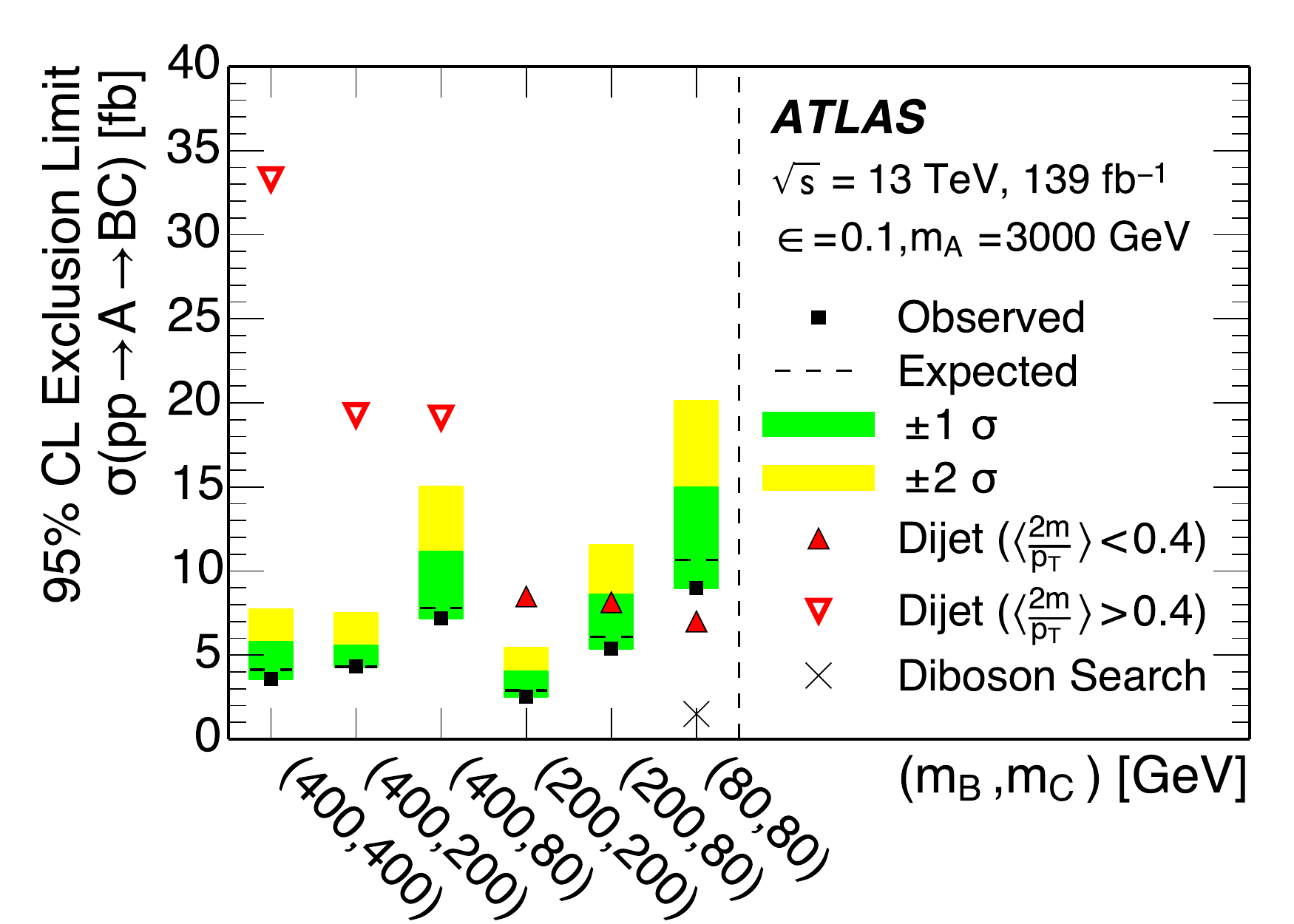}}
\subfloat[]{\includegraphics[width=0.45\textwidth]{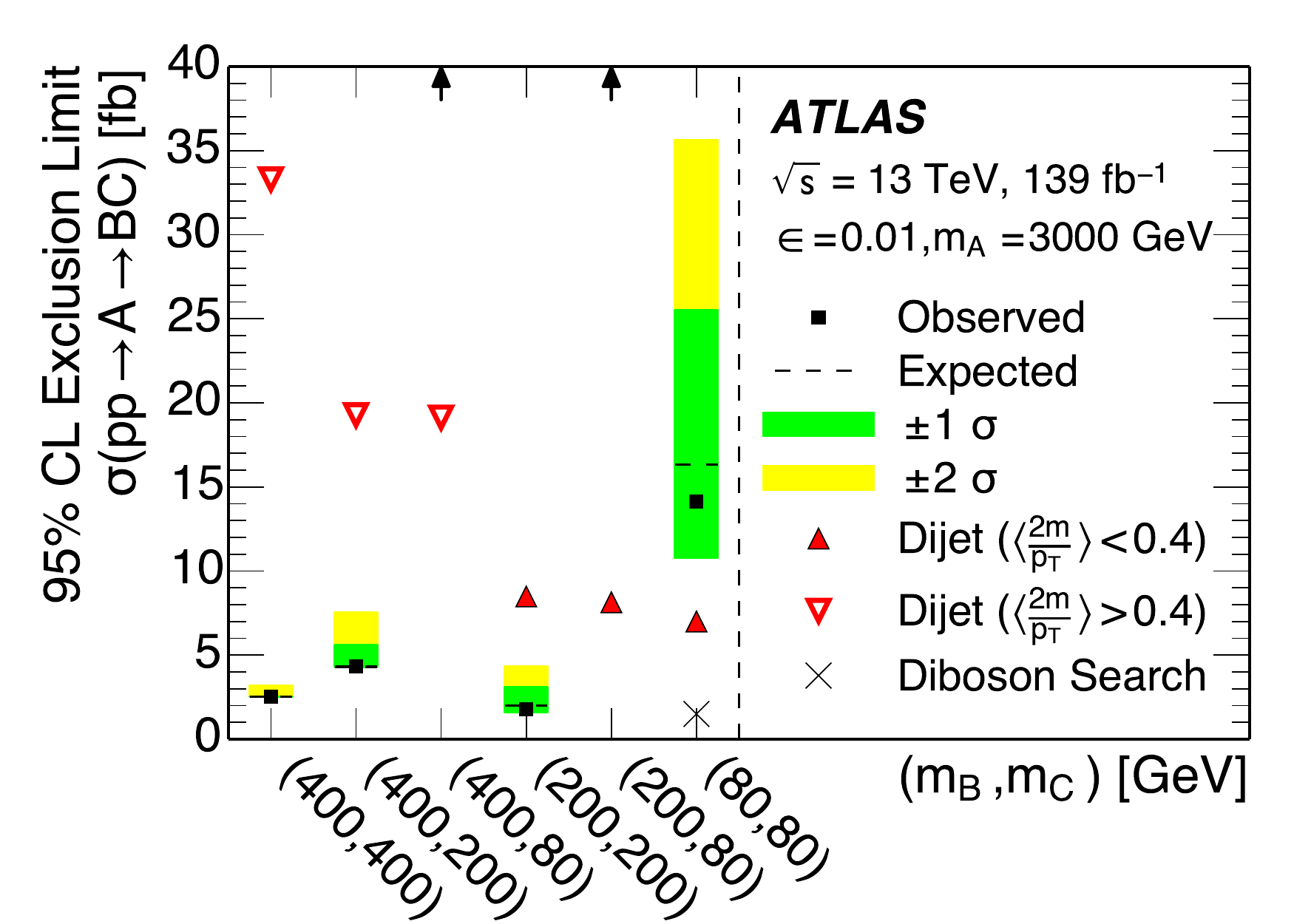}}\\
\subfloat[]{\includegraphics[width=0.45\textwidth]{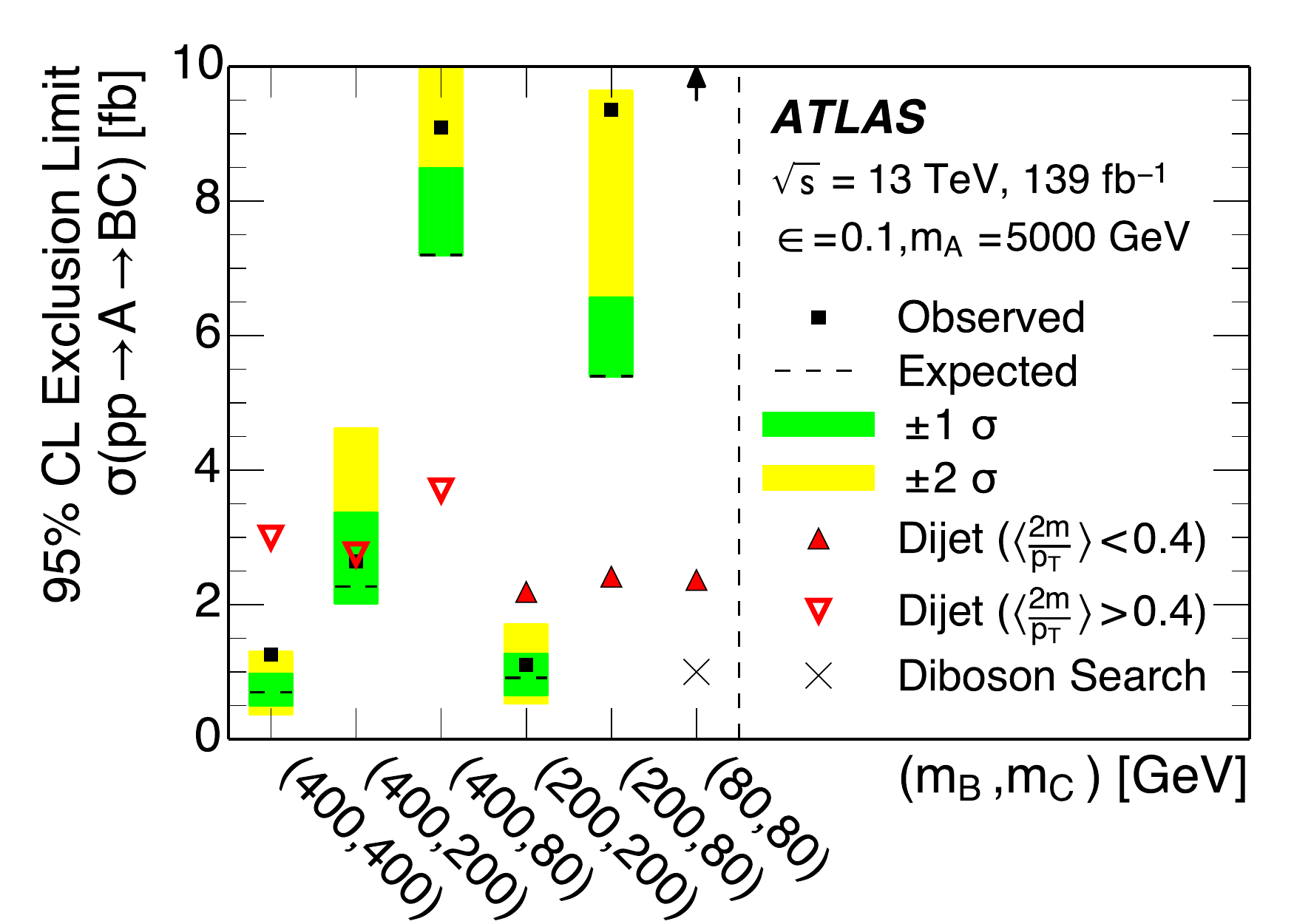}}
\subfloat[]{\includegraphics[width=0.45\textwidth]{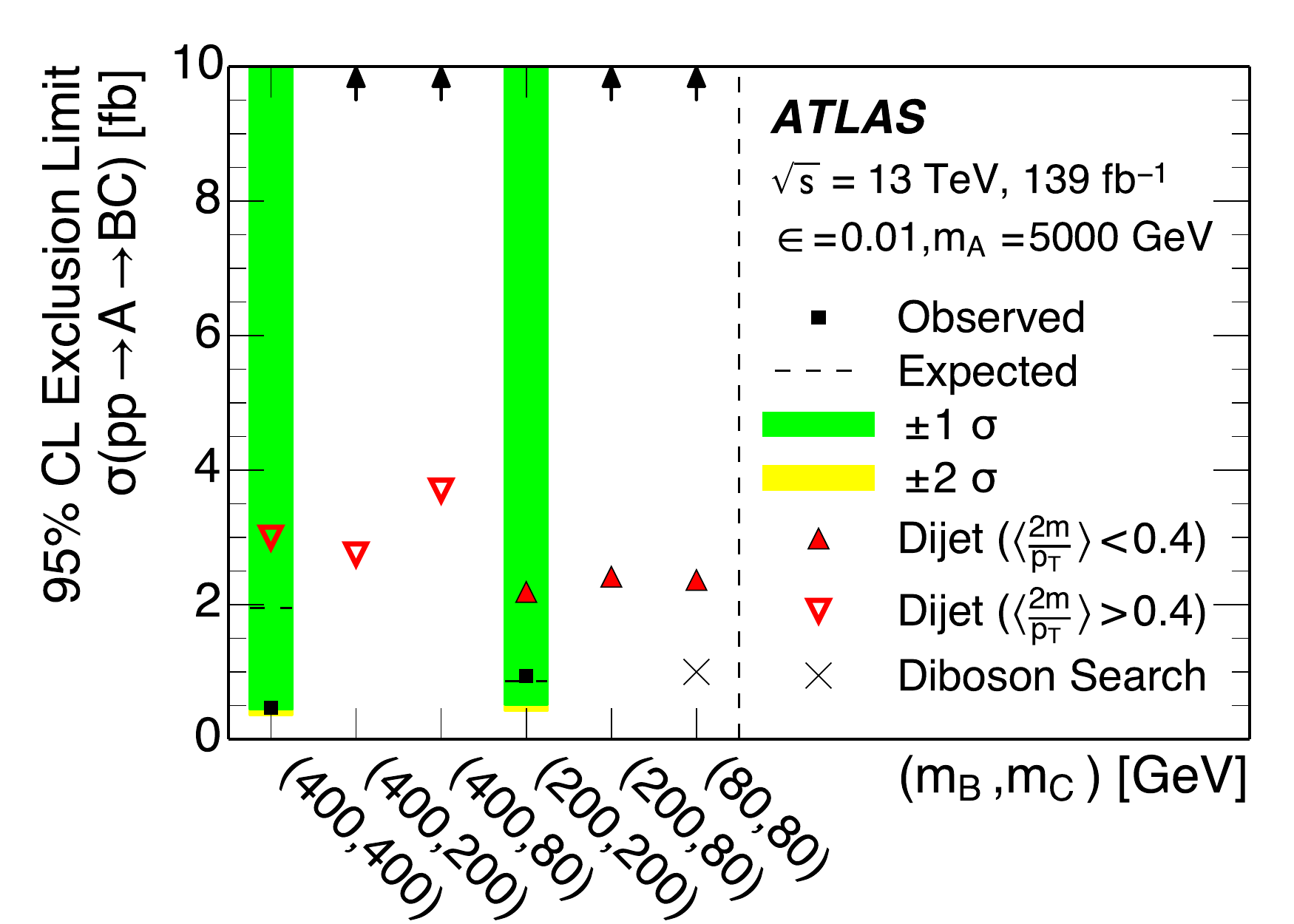}}
\caption{95\% confidence level upper limits on the cross section for a variety of signal models, labeled by $(m_B,m_C)$, in \GeV.
The limits are shown for signal models with (a,b) $m_A=3000$~\GeV\ and NN trained on signal region 2; and (c,d) $m_A=5000$~\GeV\ and NN trained on signal region 5.
The limits are broken down between the analyses with (a,c) $\epsilon=0.1$ and (b,d) $\epsilon=0.01$.  Also shown are the limits from the ATLAS dijet search~\cite{Aad:2019hjw} and the ATLAS all-hadronic diboson search~\cite{Aad:2019fbh}. The inclusive dijet limits are calculated using the $W’$ signals from this paper and the full analysis pipeline of Ref.~\cite{Aad:2019hjw}; the diboson search limits are computed using the Heavy Vector Triplet~\cite{Pappadopulo:2014qza} $W’$ signal from Ref.~\cite{Aad:2019fbh}.  The acceptance for the $W'$ in this paper, compared to the $W'$ acceptance in Ref.~\cite{Aad:2019fbh}, is 86\% and 54\% for $m_{W'}=3$ and 5~\TeV, respectively.  Missing observed markers are higher than the plotted range.  Poor limits occur when the NN fails to tag the signal.}
\label{fig:limits}
\end{figure}
 
While the regions are chosen with definite boundaries, the analysis is sensitive to signals across the entire range.  In particular, it is found that for nearly 75\% of the range, the efficiency for a signal is unaffected by a shifted peak location.  This efficiency is everywhere above half of the nominal efficiency.  
 
In conclusion, this Letter presents a model-agnostic resonance search in the all-hadronic final state using the full LHC Run 2 $pp$ data set of the ATLAS experiment.
Weakly supervised classification NNs are used to identify the presence of potential signals without training on simulations of any particular signal models.
For jets produced from Lorentz-boosted heavy-particle decays, this search is more sensitive than the inclusive dijet search and extends the coverage of the all-hadronic diboson search to regions away from the SM boson masses.
This is the first search that covers $A\rightarrow BC$ production where all of $A,B$ and $C$ are BSM particles that can have different masses.
The feature space used by the NNs is only two-dimensional, so there is great potential to extend this method to include additional features and more final states in order to ensure broad coverage of unanticipated scenarios.

\section*{Acknowledgments}
 

We thank CERN for the very successful operation of the LHC, as well as the
support staff from our institutions without whom ATLAS could not be
operated efficiently.
 
We acknowledge the support of ANPCyT, Argentina; YerPhI, Armenia; ARC, Australia; BMWFW and FWF, Austria; ANAS, Azerbaijan; SSTC, Belarus; CNPq and FAPESP, Brazil; NSERC, NRC and CFI, Canada; CERN; CONICYT, Chile; CAS, MOST and NSFC, China; COLCIENCIAS, Colombia; MSMT CR, MPO CR and VSC CR, Czech Republic; DNRF and DNSRC, Denmark; IN2P3-CNRS and CEA-DRF/IRFU, France; SRNSFG, Georgia; BMBF, HGF and MPG, Germany; GSRT, Greece; RGC and Hong Kong SAR, China; ISF and Benoziyo Center, Israel; INFN, Italy; MEXT and JSPS, Japan; CNRST, Morocco; NWO, Netherlands; RCN, Norway; MNiSW and NCN, Poland; FCT, Portugal; MNE/IFA, Romania; MES of Russia and NRC KI, Russia Federation; JINR; MESTD, Serbia; MSSR, Slovakia; ARRS and MIZ\v{S}, Slovenia; DST/NRF, South Africa; MINECO, Spain; SRC and Wallenberg Foundation, Sweden; SERI, SNSF and Cantons of Bern and Geneva, Switzerland; MOST, Taiwan; TAEK, Turkey; STFC, United Kingdom; DOE and NSF, United States of America. In addition, individual groups and members have received support from BCKDF, CANARIE, Compute Canada and CRC, Canada; ERC, ERDF, Horizon 2020, Marie Sk{\l}odowska-Curie Actions and COST, European Union; Investissements d'Avenir Labex, Investissements d'Avenir Idex and ANR, France; DFG and AvH Foundation, Germany; Herakleitos, Thales and Aristeia programmes co-financed by EU-ESF and the Greek NSRF, Greece; BSF-NSF and GIF, Israel; CERCA Programme Generalitat de Catalunya and PROMETEO Programme Generalitat Valenciana, Spain; G\"{o}ran Gustafssons Stiftelse, Sweden; The Royal Society and Leverhulme Trust, United Kingdom.
 
The crucial computing support from all WLCG partners is acknowledged gratefully, in particular from CERN, the ATLAS Tier-1 facilities at TRIUMF (Canada), NDGF (Denmark, Norway, Sweden), CC-IN2P3 (France), KIT/GridKA (Germany), INFN-CNAF (Italy), NL-T1 (Netherlands), PIC (Spain), ASGC (Taiwan), RAL (UK) and BNL (USA), the Tier-2 facilities worldwide and large non-WLCG resource providers. Major contributors of computing resources are listed in Ref.~\cite{ATL-SOFT-PUB-2020-001}.
 
 
\printbibliography
 
\clearpage 
 
\begin{flushleft}
\hypersetup{urlcolor=black}
{\Large The ATLAS Collaboration}

\bigskip

\AtlasOrcid[0000-0002-6665-4934]{G.~Aad}$^\textrm{\scriptsize 102}$,    
\AtlasOrcid[0000-0002-5888-2734]{B.~Abbott}$^\textrm{\scriptsize 128}$,    
\AtlasOrcid{D.C.~Abbott}$^\textrm{\scriptsize 103}$,    
\AtlasOrcid[0000-0002-2788-3822]{A.~Abed~Abud}$^\textrm{\scriptsize 36}$,    
\AtlasOrcid[0000-0002-1002-1652]{K.~Abeling}$^\textrm{\scriptsize 53}$,    
\AtlasOrcid[0000-0002-2987-4006]{D.K.~Abhayasinghe}$^\textrm{\scriptsize 94}$,    
\AtlasOrcid[0000-0002-8496-9294]{S.H.~Abidi}$^\textrm{\scriptsize 166}$,    
\AtlasOrcid[0000-0002-8279-9324]{O.S.~AbouZeid}$^\textrm{\scriptsize 40}$,    
\AtlasOrcid{N.L.~Abraham}$^\textrm{\scriptsize 155}$,    
\AtlasOrcid[0000-0001-5329-6640]{H.~Abramowicz}$^\textrm{\scriptsize 160}$
% ...
\bigskip
\\

% ...
$^{36}$CERN, Geneva; Switzerland.\\
% ...
$^{40}$Niels Bohr Institute, University of Copenhagen, Copenhagen; Denmark.\\
% ...
$^{53}$II. Physikalisches Institut, Georg-August-Universit\"{a}t G\"ottingen, G\"ottingen; Germany.\\
% ...
$^{94}$Department of Physics, Royal Holloway University of London, Egham; United Kingdom.\\
% ...
$^{102}$CPPM, Aix-Marseille Universit\'e, CNRS/IN2P3, Marseille; France.\\
$^{103}$Department of Physics, University of Massachusetts, Amherst MA; United States of America.\\
% ...
$^{128}$Homer L. Dodge Department of Physics and Astronomy, University of Oklahoma, Norman OK; United States of America.\\
% ...
$^{155}$Department of Physics and Astronomy, University of Sussex, Brighton; United Kingdom.\\
% ...
$^{160}$Raymond and Beverly Sackler School of Physics and Astronomy, Tel Aviv University, Tel Aviv; Israel.\\
% ...
$^{166}$Department of Physics, University of Toronto, Toronto ON; Canada.\\
% ...

\end{flushleft}


\end{document}